%Paper: gr-qc/9308005
%From: j_halliwell@icva.DNET.NASA.GOV
%Date: Fri, 6 Aug 93 10:04:48 -0400

%%%%%%%%%%%%%%%%%%%%%%%%%
%%% Paper begins here. This file is in plain Tex with no inputs required.
%%%%%%%%%%%%%%%%%%%%%%%%%
\def\refto#1{$^{#1}$}

\def\la{\langle}
\def\ra{\rangle}

\def\x{{\bar x}}
\def\y{{\bar y}}

\def\a{\alpha}
\def\b{\beta}

\def\Tr{{\twelverm Tr}}
\def\ih{{ {i \over \hbar} }}
\def\au{{\underline{\alpha}}}
\def\bu{{\underline{\beta}}}
\def\s{{\sigma}}

%%%%%%%%%%%%%%%%%%%%%%%%%%%%%
\headline={\ifnum\pageno=1\firstheadline\else
\ifodd\pageno\rightheadline \else\leftheadline\fi\fi}
\def\firstheadline{\hfil}
\def\rightheadline{\hfil}
\def\leftheadline{\hfil}
	\footline={\ifnum\pageno=1\firstfootline\else\otherfootline\fi}
\def\firstfootline{\rm\hss\folio\hss}
\def\otherfootline{\hfil}

\font\twelvebf=cmbx10 scaled\magstep 1
\font\twelverm=cmr10 scaled\magstep 1
\font\twelveit=cmti10 scaled\magstep 1

\font\tenbf=cmbx10
\font\tenrm=cmr10
\font\tenit=cmti10

\parindent=1.5pc
\hsize=6.0truein
\vsize=8.5truein
\nopagenumbers

\centerline{\tenbf ASPECTS OF THE DECOHERENT HISTORIES}
\baselineskip=22pt
\centerline{\bf APPROACH TO QUANTUM MECHANICS}
\baselineskip=16pt
\vglue 0.8cm
\centerline{\tenrm J. J. HALLIWELL}
\baselineskip=13pt
\centerline{\tenit Theory Group, Blackett Laboratory, Imperial College}
\baselineskip=12pt
\centerline{\tenit London, SW7 2BZ, United Kingdom}
\vglue 0.8cm
\centerline{\tenrm ABSTRACT}
\vglue 0.3cm
{\rightskip=3pc
 \leftskip=3pc
 \tenrm\baselineskip=12pt\noindent
I give an informal overview of the decoherent histories approach to
quantum mechanics, due to Griffiths, to Omn\`es, and to Gell-Mann and
Hartle is given. Results on the connections between decoherence,
records, correlation and entropy are described. The emphasis of the
presentation is on understanding the broader meaning of the
conditions of consistency and decoherence, and in particular, the
extent to which they permit one to assign definite properties
to the system. The quantum Brownian motion model is briefly
discussed.
\vglue 0.6cm}
\medskip
\centerline{ Imperial College Preprint 92-93/XX, August 1993}
\medskip
\noindent{\twelveit To appear in proceedings of the workshop,
``Stochastic Evolution of
Quantum States in Open Systems and Measurement Processes'',
Budapest, March 23-25, 1993 (edited by L. Di\'osi).}
\medskip
\vfil
\twelverm\baselineskip=14pt
\leftline{\twelvebf 1. Introduction}
\vglue 0.3cm
This is an informal
account of the decoherent histories
approach to quantum mechanics, loosely based on my talk at the
workshop and the questions it generated. My aim is to give a brief
summary of the approach, with some development of points not
discussed elsewhere.
Technical details will be kept to a minimum. They may be found in
the excellent articles by the subject's founders\refto{1-6}.

In a single sentence,
the broad aim of the histories approach is this:
we would like to be able to {\twelveit talk about} the properties
of a closed and isolated quantum system, without having to resort to
notions of measurement or observation in an essential way. Here, by
``talk about'', I mean make statements about the system pertaining
to its physical properties, that may be related to each other by
ordinary classical ({\twelveit i.e.}, Boolean) logic.

\vglue 0.3cm
\noindent{\twelveit 1.1 Copenhagen}
\vglue 1pt

Standard quantum mechanics
is based on the Copenhagen interpretation.  Formally, it may be
founded on a (quite large) number of technical
axioms (see, for example, Ref.[7]). These axioms place great
emphasis on the notion of {\twelveit measurement} of the system of
interest by an external, classical observing apparatus.
Indeed, the whole framework strictly applies only to
a universe which has been divided into macroscopic
classical systems and microscopic quantum systems.

Despite its great successes, it is inadequate on a number of counts.
Many -- if not all -- of these inadequacies boil down to the fact that
the Copehagen interpretation does not supply a
{\twelveit picture} of what is actually happening, or at least, what one
can  meaningfully think of as actually happening, in a closed
quantum system. It does not tell us for example, what is happening
in  a quantum system {\twelveit between} measurements. More generally, it
gives no indication as to what extent we can regard quantum systems
as possessing {\twelveit definite properties}, independently of whether or
not they are being measured. These
features of the Copenhagen interpretation make it difficult to
extend it to the macroscopic domain, and in particular,
to quantum cosmology.

\vglue 0.3cm
\noindent{\twelveit 1.2 The Histories Approach}
\vglue 1pt

The decoherent histories approach was designed to overcome these
problems. In brief, the main features of the approach are as
follows. It applies specifically to closed systems.  It focuses on
the histories of a closed system,  rather than events at a fixed
moment of time. It is a modest generalization of ordinary quantum
mechanics, but relies on a far smaller list of axioms. These axioms
are basically the statements that the closed system is described by
the usual mathematical machinery of quantum mechanics, Hilbert
space, unitary evolution of states, {\twelveit etc.}, together with a
formula for the probabilities and a rule of interpretation.  It
makes no distinction between microscopic and macroscopic. A separate
classical domain is therefore not assumed, but may be an emergent
feature under calculable conditions. It makes no essential use of
measurement, or collapse of the wave function, although these
notions may be discussed within the framework of the approach.
What replaces measurement is the more general notion of consistency
(or the stronger notion of decoherence),
determining which histories may be assigned
probabilities.  The approach also stresses classical logic, the
conditions under which it may be applied, and thus, the conditions
under which one can meaningfully talk about the properties of a
physical system.

\vglue 0.3cm
\noindent{\twelveit 1.3 Why histories?}
\vglue 1pt

The basic building block in the decoherent histories approach
is a history -- a sequence of events at a succession of times.
Why are these objects of particular interest?

\item{(a)} Histories are the most general class of situation one
might be interested in. For example, a typical experimental
situation might be of the form,
a particle is emitted from a decaying nucleus at time $t_1$, then it
passes through a magnetic field at time $t_2$, then it is absorbed
by a detector at time $t_3$.

\item{(b)} We would like to understand how classical behaviour can
emerge from the quantum mechanics of closed systems. It is therefore
necessary to show, amongst other things, that successive positions
in time of, say, a particle are approximately correlated according
to classical laws. It is therefore necessary to
study histories of position samplings.

\item{(c)} The basic practical aim of theoretical physics is to find
patterns in presently existing data. Why then should we not attempt
to formulate our theories in the terms of the density matrix of the
entire universe on a given spacelike surface? There are at least two
reasons why not. Firstly, present records are stored in a wide
variety of different ways -- in computer memories, on photographic
plates, on paper, in our own personal memories, in measuring
apparatus. A theory explaining the correlation between present
records would thus have to be a theory of measuring apparatus,
photographic plates, {\twelveit etc.} Surely a theory as fundamental as
quantum mechanics should not become so embroiled in the details of
measurement and storage. Secondly, the correlation between present
records and past events can never be perfect. In order to discuss
the approximate nature of correlations between past and present
events it becomes necessary to talk about the histories of a system.

\item{(d)} As stated above, the minimal pragmatic aim of theoretical physics
is to explain the data. Yet many feel that our theories should
explain more than just the numbers: it should supply us with a
picture of the world the way it really is. Histories arguably
supply us with that picture.

\vglue 0.6cm
\noindent {\twelvebf 2. The Formalism of Decoherent Histories}
\vglue 0.4cm
I now briefly outline the mathematical formalism of the
decoherent histories approach. Further details may be found in
Refs.[1-6,8].
\vglue 0.3cm
\noindent{\twelveit 2.1 Probabilities for Histories}
\vglue 1pt
In quantum mechanics, propositions
about the attributes of a system at a fixed moment of time are
represented by sets of projections operators. The projection
operators $P_{\a}$ effect a partition of the possible alternatives
$\a$ a system may exhibit at each moment of time. They are
exhaustive and exclusive,
$$
\sum_{\a} P_{\a} =1, \quad \quad
P_{\a} P_{\beta} = \delta_{\a \beta} \ P_{\a}
\eqno(2.1)
$$
A quantum-mechanical history is characterized by a string of
time-dependent projections,
$P_{\a_1}^1(t_1) $ $\cdots P_{\a_n}^n(t_n)$, together with an initial
state $\rho$. The time-dependent projections are related to the
time-independent ones by
$$
P^k_{\a_k}(t_k) = e^{i H(t_k-t_0)} P^k_{\a_k} e^{-i H(t_k-t_0)}
\eqno(2.2)
$$
where $H$ is the Hamiltonian.
The candidate probability for such histories is
$$
p(\a_1, \a_2, \cdots \a_n) = \Tr \left( P_{\a_n}^n(t_n)\cdots
P_{\a_1}^1(t_1)
\rho P_{\a_1}^1 (t_1) \cdots P_{\a_n}^n (t_n) \right)
\eqno(2.3)
$$
It is straightforward to show that (2.3) is both non-negative and
normalized to unity when summed over $\a_1, \cdots \a_n$. However,
(2.3) does not satisfy all the axioms of probability theory, and for
that reason it is referred to as a candidate probability. It does
not satisfy the requirement
of additivity on disjoint regions of sample space. More precisely,
for each set of histories, one may construct coarser-grained
histories by grouping the histories together. This may be achieved,
for example, by summing over the projections at each moment of time,
$$
{\bar P}_{{\bar \a}} = \sum_{\a \in {\bar \a} } P_{\a}
\eqno(2.4)
$$
although this is not the most general type of coarse graining.
The additivity requirement is then that the probabilities for each
coarser-grained history should be the sum of the probabilities of
the finer-grained histories of which it is comprised.
Quantum-mechanical interference generally prevents this requirement
from being satisfied; thus histories of closed quantum systems
cannot in general be assigned probabilities.

The standard illustrative example is the double slit experiment. The
histories consist of projections at two moments of time: projections
determining which slit the particle went through at time $t_1$, and
projections determing the point at which the particle hit the screen
at time $t_2$.
As is well-known, the probability distribution for
the interference pattern on the screen cannot be written as a sum of
the probabilities for going through each slit; hence the
candidate probabilities do not satisfy the additivity requirement.

There are, however, certain types of histories for which
interference is negligible, and the candidate probabilities for histories
do satisfy the sum rules.
These histories may be found
using the decoherence functional:
$$
D({\underline {\a}} , {\underline {\a}'} ) =
\Tr \left( P_{\a_n}^n(t_n)\cdots
P_{\a_1}^1(t_1)
\rho P_{\a_1'}^1 (t_1) \cdots P_{\a_n'}^n (t_n) \right)
\eqno(2.5)
$$
Here $ {\underline {\a}} $ denotes the string $\a_1, \a_2, \cdots
\a_n$. If the real part of
decoherence functional vanishes for all distinct pairs of
histories ${\underline {\a}}, {\underline {\a}'}$, then it may be
shown that all probability sum rules are satisfied.
Sets of such histories are said to be {\twelveit consistent}, or
{\twelveit weakly decoherent.}

A stronger condition is that both real
and imaginary parts of the decoherence functional vanish,
$$
D(\au, \au') = 0, \quad for \quad \au \ne \au'
\eqno(2.6)
$$
This I shall refer to quite simply as {\twelveit decoherence} (although it
is sometimes further classified as
{\twelveit medium} or {\twelveit strong} decoherence\refto{2}).
We will discuss this condition extensively in the next section.

The decoherence functional obeys a simple inequality which turns out
to be rather useful\refto{8}. It is,
$$
\bigl| D(\au, \au') \bigr|^2  \ \le \ D(\au,\au) \ D(\au', \au')
\eqno(2.7)
$$
Intuitively, this result says that there can be no interference
with a history which has candidate probability zero. It is useful
mathematically because if connects the off-diagonal components of
the decoherence functional to the candidate probabilities -- more
familiar objects about which more is known.
We will exploit this fact in the next section.
Eq.(2.7) also
suggests a possible measure of approximate decoherence: we say that
a system decoheres to order $\epsilon$ if the decoherence functional
satisfies (2.7) with a factor of $\epsilon^2$ on the
right-hand side. As shown in Ref.[8], such a condition implies that
most (but not all) probability sum rules will then be satisfied to
order $\epsilon$.

The focus of the decoherent histories approach is on sets of
histories satisfying the decoherence condition (2.6) (or the weaker
condition of consistency). Whether or not the decoherence condition is
satisfied will depend on the initial state, the Hamiltonian, and the
projections at each moment of time. Changing any one of these in a
decoherent set of histories will generally not preserve decoherence.
One typically expects, for a given system, to be supplied with the
initial state and the Hamiltonian. It is then matter for
investigation to determine which histories, {\twelveit i.e.}, which
strings of projections, will lead to the decoherence condition being
satisfied.

\vglue 0.3cm
\noindent{\twelveit 2.2 Consistency and Classical Logic}
\vglue 1pt

Now I discuss why sets of consistent histories are of interest.
As stated, propositions about the attributes of a quantum system may
be represented by projection operators. The set of all projections
have the mathematical structure of a lattice. This lattice is
non-distributive, and this means that the corresponding propositions
may not be submitted to Boolean logic. Similar remarks hold for
the more complex propositions expressed by general sets of
quantum-mechanical histories.

The reason why {\twelveit consistent} sets of histories are of interest is
that they {\twelveit can} be submitted to Boolean logic. Indeed, a theorem
of Omn\`es states that a set of histories forms a consistent
representation of Boolean logic if and only if it is a consistent
set\refto{6}. That is, in a consistent set of histories, each history
corresponds to a proposition about the properties of a physical
system and we can meaningfully manipulate these propositions without
contradiction using ordinary classical logic. It is in this sense
that the decoherent histories approach allows one to ``talk about''
the properties of a system in a meaningful way.

Based on these considerations, Omn\`es introduced the following rule
of interpretation of quantum mechanics alluded to in the
Introduction:
{\twelveit Any description of a physical system should consist of
propositions belonging to a common consistent quantum logic and any
reasoning about it should consist of valid implications}\refto{6}.

As an example, consider the case of retrodiction of the past, given
present data.
Suppose we have a consistent set of histories.
We would say that the alternative $\a_n$ (present data)
implies the alternatives
$\a_{n-1} \cdots \a_1 $ (past events) if
$$
p(\a_1, \cdots \a_{n-1} | \a_n ) \equiv
{ p(\a_1, \cdots \a_n )  \over p(\a_n) } = 1
\eqno(2.8)
$$
In this way, we can in quantum mechanics build a picture of
{\twelveit what actually happened} in the past, given the present data,
using only logic and the consistency of the histories. It is
not necessary for a measuring device to actually be there in the
past. Similarly, one can in certain circumstances use logic
and consistency to deduce what is happening in quantum systems
between measurements.

There is however, a caveat. Some situations in quantum mechanics
admit multiple representations of logic that are not equivalent.
They are described by two or more inequivalent sets of consistent
histories the union of which is not a consistent set. There then
exist statements about the system that are logically implied in
some sets of histories but not in all. Such statements are referred
to as ``reliable'' rather than ``true''. If, for example, the
retrodiction process outlined above suffered from
this type of ambiguity, one
would not say that the past alternatives ``actually happened''.
See Ref.[9] for a discussion of this very subtle issue.

\vglue 0.6cm
\noindent {\twelvebf 3. Decoherence, Correlation and Records}
\vglue 0.4cm

Physically, decoherence characterized by Eq.(2.6)
is intimately related to the storage of
information about a system of interest somewhere in the universe. It
is in this sense that decoherence replaces and generalizes the
notion of measurement in ordinary quantum mechanics.
Systems decohere, and hence acquire
definite properties, not necessarily through measurement, but
through their interactions and correlations with
other sytems. In this section I discuss these issues.

\vglue 0.3cm
\noindent {\twelveit 3.1 Records Imply Decoherence}
\vglue 1pt

Consider a closed system $S$ which consists of two interacting
subsystems $A$ and $B$. The Hilbert space ${\cal H}$ of $S$ is
therefore of the form ${\cal H}_A \otimes {\cal H}_B$.
Suppose we are interested in the histories characterized solely by
properties of system $A$, thus $B$ is regarded as the environment.
The system is analyzed using the decoherence functional (2.5), where
we take the $P_{\a}$ to denote a projection on ${\cal H}_A$ (strictly one
should therefore write $P_{\a} \otimes I^B$, where $I^B$ denotes
the identity on ${\cal H}_B$, but for convenience I
largely neglect this notation). I also introduce projections
$R_{\b}$ on the Hilbert space ${\cal H}_B$.

I shall show that histories of $A$ decohere if the sequences
of alternatives the histories consist of exhibit {\twelveit exact} and
{\twelveit persistent} correlations with sequences of alternatives of $B$.
More precisely,
I imagine that the alternatives of $A$ characterized by
$P_{\a_k}^k$ at each moment of time $t_k$ are perfectly recorded in
$B$ as a result of their interaction. I also imagine that this
record in $B$ is perfectly persistent ({\twelveit i.e.}, permanent).
This means that at any time $t_f$ after
the time $t_n$ of the last projection on $A$ there exist a sequence
of alternatives of $B$, $\b_1, \cdots \b_n$, that are in perfect
correlation with the alternatives of $A$, $\a_1 \cdots \a_n$ at
times $t_1 \cdots t_n $.

Correlations between subsystems are generally analyzed using the
joint probability distribution
$$
p(\a,\b) = \Tr \left( P_{\a} \otimes R_{\b} \ \rho \right)
\eqno(3.1)
$$
This is a {\twelveit bona fide} probability because it involves projections at
a
single moment of time and, by the cyclic property of the trace,
histories consisting of alternatives at a single moment of time
automatically decohere.
Then the alternatives of $A$ and $B$
characterized by the projections $P_{\a}$ and $R_{\b}$ are
said to be exactly correlated if
$$
p(\a, \b) = \delta_{\a \b} \ p(\a)
\eqno(3.2)
$$
The decoherence functional (2.6) may be written,
$$
D(\au, \au') = \sum_{\b_1 \cdots \b_n} \ \Tr \left(
R_{\b_1}^1 \cdots R_{\b_n}^n P_{\a_n}^n(t_n) \cdots P_{\a_1}^1(t_1)
\ \rho \ P_{\a_1'}^1(t_1)  \cdots P_{\a_n'}^n(t_n) \right)
\eqno(3.3)
$$
using the exhaustivity of the projections $R_{\b_k}^k$.
Now using the inequality (2.7), one finds,
$$
\bigl| D(\au, \au') \bigr| \ \le \ \sum_{\b_1 \cdots \b_n}
\left[ p(\au, \bu) p(\au', \bu) \right]^{1/2}
\eqno(3.4)
$$
where $p(\au,\bu)$ denotes the diagonal elements of the summand in
the right-hand side of (3.3).  Now the object is to show that the
strings of alternatives $\au$ and $\bu$ are perfectly correlated,
and hence the right-hand side of (3.4) vanishes unless $\au = \au'$.
Since $p(\au,\bu)$ are only candidate probabilities, we cannot
use them to discuss these correlations. However, they are
non-negative, and we may use this to derive the following
simple inequality:
$$
\eqalignno{
p(\au, \bu) \ & \le \ \sum_{\b_1 \cdots \b_{k-1} }
\ \sum_{\b_{k+1} \cdots \b_n} \ \sum_{\a_{k+1}
\cdots \a_n} \ p(\a_1, \cdots \a_n, \b_1, \cdots \b_n )
\cr & = \
\ \Tr \left( R_{\b_k}^k \otimes P_{\a_k}^k(t_k) \
\rho_{eff}(\a_1, \cdots \a_{k-1}) \right) \ p(\a_1, \cdots \a_{k-1} )
&(3.5) \cr}
$$
Here
$$
\rho_{eff}(\a_1, \cdots \a_{k-1}) =
{ P_{\a_{k-1}}^{k-1}(t_{k-1})
\cdots P_{\a_1}^1(t_1) \rho P_{\a_1}^1(t_1) \cdots
P_{\a_{k-1}}^{k-1}(t_{k-1})
\over p(\a_1, \cdots \a_{k-1} ) }
\eqno(3.6)
$$
is the effective density matrix at time $t_k$.
The first factor in the
right-hand side of (3.5) is now of the form (3.1), and
the degree of correlation between the alternatives $\a_k$ and $\b_k$
may be discussed. Note that the relevant density operator is not
$\rho$ but the effective density operator (3.6). (A further sum over
$\a_1 \cdots \a_{k-1}$ could be performed on the right-hand side of
(3.5), preserving the inequality, in which case it would be the
averaged effective density operator that would enter). In the case
of perfect correlation assumed here,
the right-hand side of (3.5) is zero, and thus $p(\au,\bu)=0$,
unless $\a_k = \b_k$. From (3.4), the decoherence
functional is therefore diagonal in $\a_k$. Carrying out the same
for all values of $k$, one finds that the decoherence functional is
diagonal in all the $\a_k$'s. This shows that, as advertized,
a perfect and persistent correlation of
alternatives of $A$ with those of $B$ leads to exact decoherence of
the histories of $A$.

This argument was inspired by an argument given by Hartle\refto{3} in his
discussion of the recovery of the Copenhagen interpretation from the
decoherent histories approach. It is, however, an improvement of his
derivation, since it utitilizes a proper definition of correlation,
(3.1), (3.2), which may be extended to the case of approximate
correlations discussed below.

\vglue 0.3cm
\noindent{\twelveit 3.2 Approximate Correlations, Entropy, Fluctuations.}
\vglue 1pt

More generally, one would expect the correlation between $A$ and $B$
to be only approximate, and the consequent decoherence will
then also be approximate. Measures of approximate correlation are
therefore required. A possible measure of the approximate
correlation in (3.1) is the mutual information,
$$
I(A;B) \ = \ \sum_{\a \b} p(\a, \b) \ln \left( p(\a, \b) \over
p(\a) p(\b) \right)
\eqno(3.7)
$$
The mutual information vanishes in the case of no correlation,
$p(\a,\b) = p(\a) p(\b)$, and is positive otherwise. A non-trivial
theorem due to Kholevo and others\refto{10} shows that the mutual information
of the joint probability distribution (3.1)
is bounded above by the von Neumann entropy of the subsystems,
$$
I(A;B) \ \le \ S[\rho_A], \quad I(A;B) \ \le \ S[\rho_B]
\eqno(3.8)
$$
Here $ S[\rho] = - \Tr ( \rho\ln \rho) $ is the von Neumann
entropy, $\rho_A = \Tr_B \rho $ is the reduced density matrix of
subsystem $A$, and similarly for $B$. This means that the von Neumann
entropy supplies an upper limit to the degree of correlation between
subsystems. From the discussion above,
it therefore also supplies an upper limit to the degree of
decoherence. (More precisely, the degree of diagonality in $\a_k$
for each $k$ is controlled by the von Neumann
entropy of the partially traced
effective density operator at time $t_k$).
This conclusion is intuitively satisfying:
decoherence is, as stated, related to the storage of information
somewhere in the universe. The degree of decoherence is therefore
limited by the physical capacity of the ``communication channel''
transmitting the information about the distinguished system to its
environment, and by the capacity of the environment to store
information.

The von Neumann entropy frequently appears in discussions of
decoherence of density matrices, where large entropy for the
distinguished subsystem is held to be  a signal of destruction of
interference\refto{11}. The above is, I believe, the first indication of a
formal connection between decoherence {\twelveit of histories} and von
Neumann entropy. It would
be of interest to explore the connection
between these notions of approximate correlation with an environment  and
the approximate decoherence condition based on Eq.(2.7).

The von Neumann entropy also appears in a related context. The
interaction with the environment typically needed for decoherence of
histories also induces fluctuations in the evolution of the
distinguished system. There is therefore a certain degree of tension
between the demands of decoherence and approximate classical
predictability: decoherence requires interaction with an
environment, which inevitably produces fluctuations, but classical
predictability requires that these fluctuations be small\refto{2}. We
will see an example of this in the quantum Brownian motion model of
the next section. In Ref.[12], an information-theoretic measure of
the size of these fluctuations was proposed -- the Shannon
information of the Husimi distribution (a certain type of smeared
Wigner function) of the density matrix $\rho$ of the distinguished
system. This measure of uncertainty is in fact bounded from below by
the von Neumann entropy of $\rho$ (Ref.[12]). This suggests that the von
Neumann entropy is the key to understanding the connection between
decoherence and fluctuations: it limits the amount of decoherence
from above but bounds the size of the fluctuations from below. Large
entropy therefore permits good decoherence, but leads to large
fluctuations; on the other hand, small entropy allows small
fluctuations, but the amount of decoherence is also small.

Much remains to be done to make these ideas precise.
They will be developed in more detail elsewhere.

\vglue 0.3cm
\noindent{\twelveit 3.3 Decoherence Implies Generalized Records}
\vglue 1pt

The decoherence achieved through persistent correlation with another
system is stronger than consistency, since both real and imaginary
parts of the decoherence functional vanish, {\twelveit i.e.}, Eq.(2.6)
holds. There is in fact a  converse to this result, namely that
Eq.(2.6) is related to the existence of records. This subsection is
a modest elaboration on the results of Gell-Mann and Hartle\refto{2}.

Consider the decoherence functional (2.5),
for any system (not just the special one discussed above).
Introduce the convenient notation
$$
C_{\au} = P_{\a_n}(t_n) \cdots P_{\a_1}(t_1)
\eqno(3.9)
$$
Let the initial state be pure, $\rho = | \Psi \ra \la \Psi |$.
In this case, the decoherence condition (2.6) is referred to as
{\twelveit medium decoherence}.
It implies that the states $ C_{\au} | \Psi \ra $
are an orthogonal (but in general incomplete) set. There therefore
exists a set of projection operators $R_{\bu}$ (not in general unique)
of which these states are eigenstates,
$$
R_{\bu} \ C_{\au} | \Psi \ra = \delta_{\au \bu} \ C_{\au} | \Psi \ra
\eqno(3.10)
$$
Note that the $C_{\au}$'s are not themselves projectors in general.
One may then consider histories consisting of the string of
projections (3.9), adjoined by the projections $R_{\bu}$ at any time
after the final time. The decoherence functional for such histories
is of the form of that in the summand in Eq.(3.3):
$$
D(\au, \bu | \au', \bu' ) = \Tr \left( R_{\bu} C_{\au} | \Psi \ra
\la \Psi | C^{\dag}_{\au'}  R_{\bu'} \right)
\eqno(3.11)
$$
These histories decohere exactly by virtue of (3.10) and (2.6),
and thus the
diagonal elements of (3.11), which we denote $p(\au, \bu)$, are true
probabilities. The correlations contained in these probabilities
may therefore be discussed. Indeed, Eq.(3.10) implies that
$p(\au,\bu) = \delta_{\au \bu} \ p(\au) $, and thus $\au$ and $\bu$
are perfectly correlated.

Medium decoherence therefore implies the existence of a string of
alternatives $\b_1 \cdots \b_n$ perfectly correlated with the string
$\a_1, \cdots \a_n$. For this reason the projection operators
$R_{\au}$ are referred to as generalized records: information about
the histories characterized by alternatives  $\a_1 \cdots \a_n$ is
recorded somewhere. It is, however, not possible to say that the
information resides in a particular subsystem, since we have not
specified the form of the system $S$; indeed, it may  not even be
possible to divide it into subsystems. It would be of interest to
study the special case in which $S$ consists of two interacting
subsystems, with the projections in (3.9) onto one of the
subsystems. One could then ask whether medium decoherence of the
distinguished subsystem alternatives implies the existence of
physical records in the other subsystem.

One can give a similar discussion of the decoherence condition (2.6)
when the initial state is mixed, although the connection with the
existence of generalized records does not appear to
come out as cleanly\refto{2}.

\vglue 0.3cm
\noindent{\twelveit 3.4 Reiteration: Consistency v. Decoherence}
\vglue 1pt

It is worth reiterating the discussions of Subsections 2.2,
3.1 and 3.3 on decoherence and consistency, and stressing their
meaning.
Consistency -- {\twelveit i.e.}, $Re D(\au, \au') = 0$ for $\au \ne \au'$
-- permits classical logic to be applied to the description of the
system. It permits one to ``talk about'' the system -- to make
statements about it which may be related to each other by classical
logic. It does not invite one to think of the system's potential
properties as actual, but it allows conditional
statements to be made: {\twelveit given}
that a certain property is actualized
one may consistently make logical deductions about which other
properties then also hold.
Decoherence -- {\twelveit i.e.} $D(\au, \au') = 0$ for $\au \ne \au'$ --
is stronger than consistency. Since it implies consistency, it also
permits classical logic to be
applied. But it is in addition related to the existence of records.
A set of histories which are decoherent, rather than merely
consistent, may be thought of as {\twelveit possessing definite
properties}, since information about those properites is stored
somewhere in the universe.

\vglue 0.6cm
\noindent {\twelvebf 4. Quantum Brownian Motion Model}
\vglue 0.4cm

I now
very briefly consider a particular model, namely the quantum
Brownian motion model. This model has been extensively studied in the
literature so only the briefest of accounts will be given here\refto{2,8,13}.
The model consists of a particle of mass $M$ in a potential $V(x)$
linearly coupled to an environment consisting of a large
bath of harmonic oscillators in a thermal state at temperature $T$.
I consider histories of position samplings of the distinguished
system. The samplings are continuous in time and Gaussian
sampling functions are used (corresponding to approximate projection
operators). The decoherence functional for the model is most
conveniently given in path-integral form:
$$
\eqalignno{
D[\x(t), \y(t)] &= \int {\cal D} x {\cal D} y \ \delta (x_f -y_f)
\ \rho(x_0, y_0)
\cr & \times \exp \left( \ih S[x(t)] - \ih S[y(t)] + \ih W[x,y] \right)
\cr & \times \exp \left(
- \int dt \ { (x(t) - \x(t) )^2 \over 2 {\s}^2 }
- \int dt \ { (y(t) - \y(t) )^2 \over 2 {\s}^2 }
\right)
&(4.1) \cr}
$$
Here, $S$ is the action for a particle in a potential $V(x)$,
$\x(t)$, $\y(t)$ are the sampled positions and $x_f$ and $x_0$
denote the final and initial values respectively.
The effects of the environment are summarized entirely by
the Feynman-Vernon
influence functional phase, $W[x,y]$, given by,
$$
\eqalignno{
W[x(t),y(t)] = & -
\int_0^t ds \int_0^s ds' [ x(s) - y(s) ] \ \eta (s-s') \ [ x(s') + y(s') ]
\cr &
+ i \int_0^t ds \int_0^s ds' [ x(s) - y(s) ] \ \nu(s-s') \ [ x(s') - y(s') ]
&(4.2) \cr }
$$
The explicit forms of the non-local kernels $\eta$ and $\nu$ may be found
in Refs.[13]. Here it is assumed, as is typical in
these models, that the initial density matrix of the total system is
simply a product of the initial system and environment density
matrices,
and the initial environment density matrix is a thermal state at
temperature $T$.
Considerable simplifications occur in a purely ohmic environment in
the Fokker-Planck limit (a particular form of the high temperature limit),
in which one has
$$
\eqalignno{
\eta(s-s') &=  M\gamma \ \delta '(s-s')
&(4.3) \cr
\nu(s-s') &= { 2 M \gamma k T \over \hbar } \ \delta (s-s')
&(4.4) \cr }
$$
where $\gamma$ is the dissipation. For convenience I will work in
this limit.

One can see almost immediately that the imaginary part of $W$,
together with the Gaussian samplings in (4.1), will
have the effect of suppressing widely differing paths $\x(t)$,
$\y(t)$. Indeed, the suppression factor will be of order
$$
\exp \left( - { 2 M \gamma k T \s^2 \over \hbar^2 } \right)
\eqno(4.5)
$$
In cgs units $\hbar \sim 10^{-27}$ and $k \sim 10^{-16}$,
so $k T / \hbar^2 \sim 10^{40} $ if $T$ is room temperature.
Values of order $1$ for $M$, $\gamma$ and $\s$ therefore
lead to an astoundingly small suppression factor.
Decoherence through interaction with a thermal
environment is thus a very effective process indeed\refto{11}.

More precisely, one can approximately evaluate the functional
integral (4.1). Let $X= (x+y)/2$, $\xi = x-y$, and use the smallness
of the suppression factor to expand about $\xi = 0$. Then the $\xi$
functional integral may be carried out with the result,
$$
\eqalignno{
D[\x(t)&,\y(t)]  =  \int {\cal D} X \ W( M \dot X_0, X_0)
\ \exp \left( - \int dt \ { ( X- {\x+\y \over 2} )^2 \over \s^2 }\right)
\cr & \times
\exp \left( - \int dt \ { F[X]^2 \over 2 (\Delta F)^2 }
- \int dt \ {(\x -\y)^2 \over 2 \ell^2 }
- i\hbar \int dt \ { (\x -\y) F[X] \over 4 \s^2 (\Delta F)^2 }
\right)
&(4.6) \cr }
$$
where
$$
F[X] = M \ddot X + M \gamma \dot X + V'(X)
\eqno(4.7)
$$
are the classical field equations with dissipation, and
$$
\eqalignno{
(\Delta F)^2 &= { \hbar ^2 \over \s^2 } + 4 M \gamma k T
&(4.8) \cr
\ell^2 &= 2 \s^2 + { \hbar^2 \over 4 M \gamma k T }
&(4.9) \cr }
$$
$W(M \dot X_0, X_0)$ is the Wigner transform of the initial density
operator.

The decoherence width (4.9) does not, in fact, immediately indicate
the expected suppression of interference, because the
temperature-dependent term will typically be utterly negligible
compared to the $\s^2$ term. The point, however, is that more precise
notions of decoherence need to be employed. One should check some of the
probability sum rules, or use the approximate decoherence condition
based on Eq.(2.7), in which the sizes of the off and
on-diagonal terms are compared. This has not been carried out for the
general expression (4.6), and in fact seems to be rather hard.
Satisfaction of the approximate
decoherence condition was checked for some special cases in Ref.[8].
Still, one expects the standard to which
decoherence is attained to be of the order of the suppression factor
(4.5), {\twelveit i.e.}, very good indeed.

Now consider the diagonal elements of the decoherence function,
representing the probabilities for histories.
$$
\eqalignno{
p[\x(t)] =  & \int {\cal D} X \ W( M \dot X_0, X_0)
\cr & \times
\ \exp \left( - \int dt \ { ( X- \x )^2 \over \s^2 }
- \int dt \ { F[X]^2 \over 2 (\Delta F)^2 }  \right)
&(4.10) \cr }
$$
The distribution is peaked about configurations $\x(t)$ satisfying
the classical field equations with dissipation; thus approximate
classical predictability is exhibited. The width of the
peak is given by (4.8). Loosely speaking, a given classical history
occurs with a weight given by the Wigner function of its initial
data. This cannot be strictly correct, because the Wigner function
is not positive in general, although it is if coarse-grained over an
$\hbar$-sized region of phase space. For a more precise discussion
of the interpretation of (4.10), see Ref.[14].

The width (2.8) has clearly identifiable contributions from quantum
and thermal fluctuations. The thermal fluctuations dominate the
quantum ones when $8 M \gamma k T \s^2 \ >> \ \hbar^2$, which, from
(4.5), is precisely the condition required for decoherence, as
previously noted\refto{12,14,15}.
Environmentally-induced fluctuations
are therefore inescapable if one is to have decoherence.

As mentioned in the previous section, there is a tension between the
demands of decoherence and classical predictability, both of which
are necessary (although generally not sufficient) for the emergence
of a quasiclassical domain\refto{2}. This tension is due to the fact that
the degree of decoherence (4.5) improves with increasing environment
temperature, but predictability deteriorates, because the
fluctuations (4.8) grow.
However,
the smallness of Boltzmann's constant ensures that  the fluctuations
(4.8) will be small compared to $F[X]$ for a wide range of temperatures
if $M$ is sufficiently large.  Moreover, the efficiency of
decoherence as evidenced through (4.5) is largely due to the
smallness of $\hbar$, and will hold for a wide range of
temperatures. So although there is some tension, there is a broad
compromise regime in which decoherence and classical predictability
can each hold extremely well.

\vglue 0.6cm
\noindent{\twelvebf 5. Concluding Remarks}
\vglue 0.4cm
In this contribution, I have given an account of the decoherent
histories approach to quantum mechanics. I have tried to cover
some aspects of the subject that are not fully described elsewhere.

For me, one of the most interesting aspects of the meeting was to
learn about the stochastic Schr\"odinger equation approaches of
various workers, including Di\'osi, Ghirardi, Gisin, Karolyhazy,
Pearle, Percival, Rimini and Spiller. Although  very different from
the decoherent histories approach, it was gratifying to discover
that some of the deepest and most difficult questions are common
to both of these approaches to quantum mechanics. An example is the
general question of the most natural
most way to divide a sufficently
large complex system into distinguished system and environment.
I feel that workers in these fields have much learn
from each other. Indeed, as a consequence of the meeting, a project
was commenced with the aim of exploring the seemingly close
connections between the quantum state diffusion approach of Gisin
and Percival\refto{16} and the decoherent histories approach\refto{17}.

For further literature on the decoherent histories approach, see
Refs.[18-28].

\vglue 0.6cm
\noindent{\twelvebf 6. Acknowledgments}
\vglue 0.4cm
I am
extremely grateful to Lajos Di\'osi for organizing such a
stimulating meeting. I would also like to thank Chris Isham,
Lajos Di\'osi,
Nicholas Gisin, Phillip Pearle,  Ian Percival, Simon Saunders, Tim
Spiller, Euan Squires, Dieter Zeh for useful conversations.
This work was supported by a University Research Fellowship from the
Royal Society.

\vglue 0.6cm
\noindent{\twelvebf References \hfil}
\vglue 0.4cm

\medskip

\itemitem{1.} M. Gell-Mann and J. B. Hartle, in {\twelveit Complexity, Entropy
and the Physics of Information, SFI Studies in the Sciences of Complexity},
Vol. VIII, W. Zurek (ed.) (Addison Wesley, Reading, 1990).

\itemitem{2.} M. Gell-Mann and J. B. Hartle,
{\twelveit Phys.Rev.} {\twelvebf D47}, 3345 (1993).

\itemitem{3.} J. B. Hartle, in {\twelveit Quantum Cosmology and Baby
Universes}, S. Coleman, J. Hartle, T. Piran and S. Weinberg (eds.)
(World Scientific, Singapore, 1991).

\itemitem{4.} J. B. Hartle, in Proceedings of the 1992 Les Houches Summer
School, {\twelveit Gravitation et Quantifications}, B.Julia (ed.).

\itemitem{5.} R. B. Griffiths, {\twelveit J.Stat.Phys.} {\twelvebf 36}, 219
(1984);
{\twelveit Phys.Rev.Lett.} {\twelvebf 70}, 2201 (1993).

\itemitem{6.} R. Omn\`es, {\twelveit Ann.Phys.} {\twelvebf 201}, 354 (1990);
{\twelveit Rev.Mod.Phys.} {\twelvebf 64}, 339 (1992).

\itemitem{7.} B. d'Espagnat, {\twelveit Conceptual Foundations of Quantum
Mechanics} (Benjamin, \ \ \ Reading, MA, 1977).

\itemitem{8.} H. F. Dowker and J. J. Halliwell, {\twelveit Phys. Rev.}
{\twelvebf
D46}, 1580 (1992).

\itemitem{9.} R. Omn\`es. {\twelveit J.Stat.Phys.} {\twelvebf 62}, 841 (1991).

\itemitem{10.} A. S. Kholevo, {\twelveit Problemy Peredachi Informatsii}
{\twelvebf
9}, 3 (1973); B. Schumacher, University of Texas PhD Thesis (1990);
H. Everett, in B. S. DeWitt and N. Graham, (eds.), {\twelveit The Many Worlds
Interpretation of Quantum Mechanics} (Princeton University Press,
Princeton, 1973).

\itemitem{11.} E. Joos and H. D. Zeh, {\twelveit Zeit.Phys.} {\twelvebf B59},
223 (1985);
W. G. Unruh and W. Zurek, {\twelveit Phys.Rev.} {\twelvebf D40}, 1071 (1989);
W. Zurek, {\twelveit Physics Today} {\twelvebf 40}, 36 (1991);
J. P. Paz and W. H. Zurek,
``Environment-Induced Decoherence, Classicality and Consistency
of Quantum Histories'', Los Alamos preprint (1993).

\itemitem{12.} A. Anderson and J. J. Halliwell,
``Information-Theoretic Measure of Uncertainty Due to Quantum and
Thermal Fluctuations'', Imperial College Preprint (1993), to appear in
{\twelveit Physical Review D}.

\itemitem{13.} R. P. Feynman and J. R. Vernon, {\twelveit Ann.Phys.(N.Y.)}
{\twelvebf 24}, 118 (1963);
A. O. Caldeira and A. J. Leggett, {\twelveit Physica} {\twelvebf 121A}, 587
(1983);
B. L. Hu, J. P. Paz and Y. Zhang, {\twelveit Phys.Rev.} {\twelvebf
D45}, 2843 (1992).

\itemitem{14.} J. J. Halliwell, ``Quantum-Mechanical Histories and
the Uncertainty Principle. I. Information-Theoretic Inequalities'',
Imperial College Preprint (1993), to appear in {\twelveit Physical Review
D};
``Quantum-Mechanical Histories and the Uncertainty Principle. II.
Fluctuations about Classical Predictability'', Imperial College
preprint (1993).

\itemitem{15.} B-L. Hu and Y.Zhang, ``Uncertainty Relation at Finite
Temperature'', University of Maryland preprint (1992).

\itemitem{16.} N. Gisin and I. C. Percival, {\twelveit J.Phys.A}, {\twelvebf
25}, 5677 (1992); {\twelvebf 26}, 2233 (1993); {\twelvebf 26}, 2245 (1993).

\itemitem{17.} L. Di\'osi, N. Gisin, J. J. Halliwell and I.
Percival, work in progress (1993).

\itemitem{18.} A. Albrecht, {\twelveit Phys.Rev.} {\twelvebf D46}, 5504 (1092).

\itemitem{19.} M. Blencowe, {\twelveit Ann.Phys.} {\twelvebf 211}, 87 (1991).

\itemitem{20.} T. Brun, {\twelveit Phys.Rev.} {\twelvebf D47}, 3383 (1993).

\itemitem{21.} E. Calzetta and B. L. Hu, ``Decoherence of
Correlation Histories'', in {\twelveit Directions in General
Relativity}, edited by B. L. Hu and T. A. Jacobson (Cambridge
University Press, Cambridge, 1993).

\itemitem{22.} L. Di\'osi,
``Unique Family of Consistent Histories in the Caldeira-Leggett
Model'',
Budapest preprint KFKI-RMKI-23 (1993).

\itemitem{23.} L. Di\'osi,
``Unique Quantum Paths by Continuous Diagonalization of the
Density Operator'',
Budapest preprint KFKI-RMKI-28 (1993).

\itemitem{24.} J. Finkelstein,
``On the Definition of Decoherence'',
San Jos\'e preprint SJSU TP-93-10 (1993).

\itemitem{25.} C. Isham,
``Quantum Logic and the Histories Approach to Quantum Theory'',
Imperial Preprint TP/92-92/39 (1993).

\itemitem{26.} R. Laflamme and A. Matacz,
``Decoherence Functional and Inhomogeneities in the Early Universe'',
Los Alamos preprint (1993).

\itemitem{27.} Y. Ohkuwa,
``Decoherence Functional and Probability Interpretation'',
Santa Barbara preprint, UCSBTH-92-40 (1992).

\itemitem{28.} J. Twamley,
``Phase Space Decoherence: A Comparison between Consistent
Histories and Environment-Induced Superselection'',
Adelaide preprint ADP-93-208/M19 (1993).

\itemitem{29.} S. Saunders,
``The Quantum Block Universe'', Harvard
Philosophy Dept. preprint (1992).

\end

\vfil\supereject
\bye